==================

\documentstyle[12pt]{article}

\begin{document}

\begin{titlepage}

\begin{flushright}
October 1993
\end{flushright}
\begin{center}

\vspace{.5 true  cm}

\Large
{\bf Non-Fermi Liquid Behaviour of Electrons }\\
\vspace{6pt}
{\bf in the Half-filled Honeycomb Lattice}\\
\vspace{6pt}
(A renormalization group approach)

\vspace{1.5 true  cm}

\large
{\bf J. Gonz\'alez\S\footnote{e-mail: emgonzalez@iem.csic.es},
F. Guinea\dag \footnote{e-mail: paco@ccuam3.sdi.uam.es} and
M.A.H. Vozmediano\ddag }

\vspace{0.5 true  cm}

\normalsize
\S{\em Instituto de Estructura de la Materia, CSIC} \\
{\em Serrano 123, 28006 Madrid, Spain} \\

\dag{\em Instituto de Ciencia de Materiales, CSIC} \\
{\em Cantoblanco, 28049 Madrid, Spain.} \\

\ddag{\em Departamento de Ingenier\'{\i}a, Universidad Carlos
III de Madrid} \\
{\em Avda. Mediterr\'aneo s/n, 28913 Legan\'es (Madrid), Spain.}

\vspace{.5 true  cm}

\end{center}

\begin{abstract}

A system of electrons in the two-dimensional
honeycomb lattice
with Coulomb interactions is described by a
renormalizable quantum field theory similar but not equal to
QED$_3$. Renormalization group techniques are used to
investigate the infrared behavior of the system that
flows to a fixed point with non-Fermi liquid
characteristics. There are anomalous dimensions in
the fermionic observables, no quasiparticle pole, and
anomalous screening of
the Coulomb interaction.
These results are robust as the Fermi level is not changed by
the interaction.
The system resembles in the infrared the one-dimensional
Luttinger liquid.

\vspace{3 true cm}

\end{abstract}

\vspace{3 true cm}
\vskip5cm
\noindent

\newpage

\end{titlepage}
\section{Introduction}

This paper addresses the problem of interacting electrons
in a band of a two-dimensional
crystal. This problem uncovers all the macroscopic
properties  of the materials of which the maximum interest
is centered at the present on the high-$T_c$ superconductors,
in particular, on the nature of their normal state.
To fix ideas and some notation let us
review the highlights of the problem.
Start with a free theory of, say, electrons  in a metal,
i. e., a free Fermi gas whose ground state is known and
whose low energy
elementary  excitations are electrons and holes obeying
a free dispersion relation.
Then switch on an interaction among the electrons and ask
the question of what the new ground state is and
what is the nature of  the low  energy excitations of the full
interacting theory. This program has been successfully addressed
by Landau in three spatial dimensions $d$ where
it leads to the  famous theory of Landau's Fermi
liquid \cite{landau} and
the concept of  quasiparticles. Things are different in
$d=1$ where an exact solution can be found and the behavior
of the system is always of the Luttinger type\cite{haldane}.

The case $d = 2$ is physically very interesting both
in practice (most of the high-$T_c$ materials are organized
in two dimensional layers and experimental data are available)
and under a theoretical point
of view. Despite the great effort devoted to it,
no exact solution is known at the present and
the situation remains controversial\footnote{One
of the  clearest examples  of
the mentioned difficulties is the ``infrared catastrophe" \cite{IR}
discovered by Anderson that is attracting a major  interest in
relation with the study of the high $T_c$ superconductivity.}.
The ultimate origin of the
trouble lies on the very complex nature of the many-body
vacuum and the corresponding  difficulties of doing perturbations
around the Fermi surface.

The theory of the
Fermi liquid and the perturbations of it
that lead to the BCS and charge density  wave
instabilities
have recently been rephrased as
effective field theories \cite{pol,gal,shan} to which
renormalization group techniques can be applied \cite{RG}.
The main idea of this approach is that, although we are dealing
with problems that naturally have an ultraviolet
cutoff $\rho_0$ (the width of the band),
we are only interested in the physics at a scale much lower
than it, namely, a very narrow strip above and below the
Fermi surface. By integrating out in the path integral all the
momenta above a certain cutoff $\rho\leq\rho_0$,
we end up with an effective field theory whose couplings
are cutoff dependent. The renormalization group transformations
are rescalings of $\rho$. In this way one can visualize
the Fermi liquid as a fixed point of this RG transformations
(it is gapless) that flows towards a BCS state when a weak
attraction is introduced between the quasiparticles.

When comparing the  effective field theories that arise in this
context with standard quantum field theory,
we encounter two main
differences. The first one  is  the nonrelativistic nature
of the many body physics where there is  a
parameter, the Fermi velocity, whose typical value is much smaller
than the speed of light. This prevents the use of the general
theorems concerning renormalizability.
The more severe difference lies
in the above mentioned nature of the many body vacuum.
When treating the interacting system as a perturbation of the
``free"  system constituted  by the fixed point, perturbations
are to be described  around a Fermi surface in $d=3$, a Fermi line
in $d=2$, and Fermi points in $d=1$. We are then faced with
couplings that are functions of the shape of the Fermi surface which is
also modified by the  interaction. Only in the $d=1$ case we  deal,
as in quantum field theory, with a true vacuum around which
one can expand and get a bunch of coupling {\it constants}.

In ref. \cite{nos}, we developed  a model for  the electronic structure of
free electrons in a two-dimensional honeycomb lattice.
This lattice constitutes the basic
building block of periodic systems made of identical atoms
with planar, threefold coordination. That is the case of
graphite, built of carbon atoms. The rapidly growing
family of fullerene
compounds \cite{sciam,fis} can be thought of as closed, curved
surfaces derived from the planar honeycomb lattice.
These compounds show a variety of unexpected
features, the most striking of them being the
superconductivity of doped crystals of C$_{60}$
molecules \cite{super1,super2}. Moreover, this superconductivity can
be considered a form of high-T$_c$ superconductivity,
as the number of electrons available to form Cooper
pairs is unusually low.

The main feature that makes the honeycomb lattice
the object of our attention in the present context is
that its Fermi level at half-filling
consists  of  exactly two Fermi {\it points}.
This peculiarity opens the possibility
of performing a complete analysis of the problems mentioned above
in analogy with the
$d=1$ case. In particular it
makes of this lattice an ideal laboratory for investigating
departures from Fermi liquid behavior in $d=2$ which have
shown very ellusive in more conventional approaches. The
major interest of this investigation relies  on the fact,
made clear by now  from the experiments, that
the normal state of high-$T_c$
materials is not of  Fermi liquid type.
Besides, it can be directly applied to the
physics of the fullerenes that, as mentioned above, are
two-dimensional
graphitic structures at half-filling \cite{app}.

The paper is organized as follows. In section 2 we review
results on the free electrons on
the honeycomb lattice that lead to the low energy effective
action used as a starting point of
our analysis. We emphasize the point that, unlike what
usually occurs in many body physics, our effective
action constitutes a genuine quantum field theory with
a single, well-defined, vacuum state. Next the
free theory is completed with the addition of the Coulomb
interaction. In the field theory, this is
mediated by an electromagnetic field
whose propagator reproduces, in the nonrelativistic
limit, the usual four-fermion interaction
considered in the many-body approach.
We end up with a standard local quantum field theory.
A discussion is done on the differences to be expected
by the use of the relativistic versus the instantaneous
electromagnetic field propagator
illustrated by the computation of the
density of states in both cases. The results announce
an infrared instability which is clarified in section 4 with the
help of the renormalization group.

Section 3 is devoted to the study of the renormalizability
of the model. It is shown that, despite its nonrelativistic
nature and due to its gauge invariance, it is renormalizable.
The Fermi velocity and the electron wave function get
renormalized while the electric charge does not.

Sections 4 and 5 contain the analysis of the
RG  flow of the model and its  physical implications.
The first section is devoted to the
nonrelativistic limit of the theory.
We see that the perturbative expansion has an effective coupling given by
the ratio of the square of the
electric charge and the Fermi velocity.
The RG shows that this  coupling
decreases to zero in the infrared although there is no
fixed point in the nonrelativistic regime. We then investigate the
possibility of having a natural length scale in the theory and
make some comments on the system away from half-filling. Finally
we find an anomalous dimension for the electron wave function
at two-loop order that points towards a non-Fermi liquid
behavior of the system.

The previous analysis is completed in
section 5 where we study the relativistic
regime which is the relevant regime in the infrared.
We find a fixed
point of the RG flow obtained when the Fermi velocity
equals the speed  of light. Next we compare our model with
standard RG analysis of the Fermi liquid theory and find
significant differences both from the technical
and from the physical point of view. The most
significant one is
the anomalous dimension found for the
electron Green function.
We discuss the non renormalization of the electric charge and
establish the non-Fermi liquid nature of our fixed point.

We set our main conclusions and a summary of the more
relevant points of the article in section 6.

\section{Many-body theory of the 2D layer}

In this section we first review the one-particle electronic
states in the two-dimensional honeycomb lattice, which serve
afterwards as a starting point for the many-body problem. The
present one-particle description has been worked out with
complete detail in ref. \cite{nos}, regarding the
conduction band of a two-dimensional graphite layer.
The honeycomb lattice has two different atoms
per primitive cell, so that
two degrees of freedom arise
in a variational computation
of the energy eigenstates,
corresponding to the
respective orbitals for the two atoms.
In the tight-binding approximation the problem reduces to the
diagonalization of the one-particle hamiltonian
\begin{equation}
H = - t \sum_{<i,j>} a_{i}^{+}a_{j}  \label{6}
\end{equation}
where the sum is over pairs of nearest neighbors atoms $i,j$ on the
lattice and $a_{i}$ , $a_{j}^{+}$ are canonically anticommuting
operators
\begin{equation}
\{ a_{i},a_{j} \} = \{ a_{i}^{+}, a_{j}^{+} \}
 = 0  \;\;\;\;\;\:\;\;\; \{ a_{i}, a_{j}^{+} \} = \delta_{ij}
\label{7}
\end{equation}
The states which are simultaneously eigenvectors of the
hamiltonian and of the lattice generators have to be of the form
\begin{equation}
\Psi = \sum_{i \;\bullet} c_{\bullet}
\mbox{\Large $e^{i{\bf k \cdot r}_{i}}$} a_{i}^{+} \left| O \right\rangle
+  \sum_{i \;\circ} c_{\circ}
\mbox{\Large $e^{i{\bf k \cdot r}_{i}}$}a_{i}^{+} \left| O \right\rangle
\label{8}
\end{equation}
with coefficients $c_{\bullet}$ and $c_{\circ}$
for black and blank points, respectively, as depicted in figure
1. Obviously, under the action of (\ref{6}) black points are mapped into
blank points and viceversa. We have, indeed,
\begin{eqnarray}
  H \Psi  & = &  - t \sum_{i \;\bullet}
\sum_{<i,j>} c_{\circ}\mbox{\Large $e^{i
{\bf k \cdot r}_{j}}$} a_{i}^{+} \left| O \right\rangle
\; - \;  t \sum_{i \;\circ} \sum_{<i,j>}c_{\bullet}
\mbox{\Large $e^{i
{\bf k \cdot r}_{j}}$} a_{i}^{+} \left| O \right\rangle
  \nonumber    \\
 & = &   - t \sum_{j}  \mbox{\Large $e^{i {\bf k \cdot u}_{j}}$}
\sum_{i \;\bullet} c_{\circ}
\mbox{\Large $e^{i {\bf k \cdot r}_{i}}$}
            a_{i}^{+} \left| O \right\rangle
\;  - \;  t \sum_{j}  \mbox{\Large $e^{i {\bf k \cdot v}_{j}}$}
\sum_{i \;\circ}  c_{\bullet}
        \mbox{\Large $e^{i {\bf k \cdot r}_{i}}$}
           a_{i}^{+} \left| O \right\rangle
\end{eqnarray}
where $\{ {\bf u}_{j} \}$ is a triad of vectors connecting a
$\bullet$ atom with its nearest neighbors, and $\{ {\bf v}_{j} \}$
the triad made of their respective opposites (see figure 1).
Therefore, the state (\ref{8}) is an eigenvector of
$H$ provided that the coefficients $c_{\bullet}$ and $c_{\circ}$
are solutions of the eigenvalue problem
\begin{equation}
     \left(
\begin{array}{cc}
 0  &  - t \sum_{j}\mbox{\Large $e^{i{\bf k \cdot u}_{j}}$} \\
- t \sum_{j}\mbox{\Large $e^{i{\bf k \cdot v}_{j}}$} &  0
\end{array} \right) \left(
\begin{array}{c}
c_{\bullet}  \\
c_{\circ}
\end{array}    \right) = E({\bf k})
\left(
\begin{array}{c}
c_{\bullet}  \\
c_{\circ}
\end{array}    \right) \label{9}
\end{equation}
The parameter $t$ measures the hopping between nearest neighbors
and provides the natural energy scale in the one-particle
description. The diagonalization of (\ref{9}) gives the band of
levels
\begin{equation}
E({\bf k}) =  \pm t \sqrt{1 + 4
cos^{2}\frac{\sqrt{3}}{2} a k_{x} +
4 cos \frac{\sqrt{3}}{2} a k_{x} \; cos \frac{3}{2}
a k_{y} }          \label{12}
\end{equation}
the parameter $a$ being the lattice spacing.
We will be interested in the consideration of this band at
half-filling, which is the pertinent instance for the
carbon-based materials and, in particular,
graphite layers and undoped fullerenes. The dispersion relation
(\ref{12}) has the amazing property that the Fermi level at
$E({\bf k}) = 0$  is only reached by six isolated points in
momentum space (see figure 2).
These are the six corners of the first Brillouin
zone and, due to the periodicity of the dual lattice,
they correspond to only two independent states.
The existence of a finite number of Fermi points is quite
unusual in more than one spatial dimension. It has important
consequences since it makes possible to encode the low-lying
excitations about the Fermi level into a simple field theory.
When considering an arbitrarily large sample, it is appropriate
to amplify the region about any of the two independent Fermi
points by taking the limit $a \rightarrow 0$. In either case we
obtain for the operator ${\cal H}$ in equation (\ref{9})
\begin{equation}
{\cal H} =  \frac{3}{2} t a \;
\mbox{\boldmath $\sigma$} \mbox{\boldmath $\cdot \delta k$}
+ O\left( (a \: \delta k )^{2} \right) \label{deq}
\end{equation}
where \mbox{\boldmath $ \delta k$} is the displacement about the
Fermi point and $\sigma_{1} , \sigma_{2}$ stand for the two
Pauli matrices. The conclusion is that, in the long-wavelength
limit $a \rightarrow 0$, the one-particle (energy) eigenstates
of the electrons in the honeycomb lattice are given by the field
theory of two massless Dirac
spinors (one for each independent Fermi point) in two spatial
dimensions\footnote{Similar dispersion relations can be found in higher
dimensions. The best known of them is the one derived from the
3D diamond structure with one state per site\cite{tosatti}.}.
This is all about the free theory.

We now turn to introduce the electronic interaction by applying
the method of second quantization. We remark that, proceeding in
this way with the above set of one-particle states, we end up
with a formalism which falls more into the framework of quantum
field theory than into that of quantum statistical theory. The
difference is subtle but significant. In quantum statistical
theory the ground state of the system is, in general, a condensate with an
extended Fermi surface (in more than one spatial dimension)
and the elementary excitations (quasiparticles) are only stable
right at the Fermi level.
The second quantized
theory built from the dispersion relation (\ref{deq}) has,
however, a definite vacuum in the sense of quantum field theory.
Its excitations are similar to the particle-hole excitations of
the Dirac sea. We may consider separately, for instance, each of
the two Dirac spinors and work out the many-body electron
propagator of the free theory
\begin{equation}
  G^{(0)}(t,\mbox{\bf r};t',\mbox{\bf r}')  =  - i \langle T \Psi_{I}(
t,\mbox{\bf r}) \overline{\Psi}_{I}(t',\mbox{\bf r}')  \rangle
\end{equation}
We denote by $\Psi_{I}$ the electron operator $\Psi$ in the
interaction representation while $\overline{\Psi}_{I} = \Psi_{I}
^{\dag} \: \sigma_{3}$. In terms of the respective modes $\psi_{{\bf
k},+} , \psi_{{\bf k},-}$ for unoccupied and occupied levels we
have\footnote{From now on we replace \mbox{\boldmath $\delta k$}
by the wavevector \mbox{\boldmath $k$} with origin at the Fermi point.}
\begin{equation}
\Psi_{I} (t, {\bf r}) = \sum_{\bf k} \mbox{\Large $e^{i {\bf k
\cdot r} } e^{- i v\left|{\bf k}\right| t}$}  \psi_{{\bf k},+}
a_{{\bf k},+} + \sum_{\bf k} \mbox{\Large $e^{i {\bf k
\cdot r} } e^{ i v\left|{\bf k}\right| t}$}  \psi_{{\bf k},-}
a_{{\bf k},-}
\end{equation}
where $v = 3 ta/2$ is the Fermi velocity
and the operators $a_{{\bf k},\pm}$ satisfy
anticommutation relations
\begin{equation}
\{ a_{{\bf k},\pm}, a_{{\bf k}',\pm}^{+} \} = \delta ({\bf k} -
{\bf k}')
\end{equation}
Taking the statistical average with the system at half-filling,
we get for the Fourier transform of the propagator
\begin{equation}
  G^{(0)}(\omega, \mbox{\boldmath $k$})  =  i \frac{ - \gamma_{0} \omega
+ v \; \mbox{\boldmath $\gamma \cdot k$}}
{ - \omega^{2} + v^{2} \mbox{\boldmath $k$}^{2} - i \epsilon } \label{prop}
\end{equation}
$ \left\{ \gamma_0 , \mbox{\boldmath $\gamma $} \right\} $
is a standard set of $\gamma $-matrices satisfying
\begin{eqnarray}
\left\{ \gamma_0 , \gamma_0 \right\} = - 2 &
   \left\{ \gamma_{i} , \gamma_{j} \right\} = 2 \delta_{ij} &
       \left\{ \gamma_0 , \gamma_{i} \right\} = 0
\end{eqnarray}
and related to the above quoted Pauli matrices by $\sigma_3 = - i
\gamma_0, \; \sigma_1 = i \sigma_3 \gamma_1, \;
 \sigma_2 = i \sigma_3 \gamma_2$.
The above expression coincides precisely with the Feynman
propagator for a massless spinor in 2 + 1 dimensions, which
confirms our assertion that the system at half-filling is
equivalent to the Dirac sea in quantum field theory.
We must notice however the presence of the Fermi velocity $v$ in
front of the spatial part of the scalar product, which is a
feature of the many-body theory.

In the nonrelativistic approximation, the introduction of the electronic
interaction leads, in first instance, to the second quantized
hamiltonian
\begin{eqnarray}
  H_{Coulomb} & = &
     \frac{3}{2} t a \int d^{2} r \; \overline{\Psi}(\mbox{\bf r})
\mbox{\boldmath $\gamma \cdot \nabla$} \Psi(\mbox{\bf r})    \nonumber  \\
 &  &  + \frac{e^{2}}{2} \int d^{2} r_{1} \int d^{2} r_{2}
\frac{\overline{\Psi}(\mbox{\bf r}_{1}) \sigma_{3}
\Psi(\mbox{\bf r}_{1})
\overline{\Psi}(\mbox{\bf r}_{2}) \sigma_{3}
\Psi(\mbox{\bf r}_{2})}{4\pi \left| \mbox{\bf r}_{1} -
 \mbox{\bf r}_{2} \right|}    \label{hcoul}
\end{eqnarray}
We attach to this hamiltonian the name of ``Coulomb'' since it
describes the instantaneous interaction between electric
charges. The model given in terms of the Coulomb interaction is,
however, a highly nonlocal field theory, what makes very
awkward its investigation from a formal point of view.
Both for the sake of studying the properties of the
quantum theory as well as for computational purposes, a complete
description of the interaction with the electromagnetic field $A_{\mu}$
is more desirable. One of the points that we support in what
follows is that, actually, the model of interacting electrons is
sensitive to retardation effects of the electromagnetic propagation.
The reason for such unconventional behaviour is the massless
character of the spinor $\Psi $. This property is not accidental
in the description of the two-dimensional
layer, as long as the expansion
(\ref{deq}) gives only one marginal operator dictating the
long-distance behaviour of the
free theory.

We propose, therefore, a quantum field theory description based
on the second quantized hamiltonian\footnote{Unless otherwise
stated, we work henceforth in units $\hbar = c = 1$.}
\begin{equation}
  H  =  \frac{3}{2} t a \int d^{2} r \; \overline{\Psi}(\mbox{\bf r})
\mbox{\boldmath $\gamma \cdot \nabla$} \Psi(\mbox{\bf r})  -  e \int
d^{2} r \; j_{\mu} A^{\mu}   \label{hq}
\end{equation}
The interaction of the electromagnetic field $A_{\mu} $ and the
electrons in the layer is described in the standard fashion,
by coupling to the conserved current
\begin{equation}
j_{\mu } \sim \left( i \: \overline{\Psi} \gamma_0 \Psi ,
  \: i \: v \overline{\Psi} \mbox{\boldmath $\gamma $} \Psi  \right)
\end{equation}
This poses some technical problems since the electromagnetic
field propagates in three-dimensional space while we want the
dynamics of the electrons to be confined to the two-dimensional layer.
Although this may not be achieved in general, it turns out to be
possible by specializing to the Feynman gauge\cite{ramond}, which enforces
the constraint
\begin{equation}
\nabla_{\mu} A^{\mu} = 0
\end{equation}
and has also the property of placing the $A_{\mu }$ field in the
same direction as the electronic current. In this gauge we have
the propagator
\begin{equation}
\langle T A_{\mu}(t,\mbox{\bf r}) A_{\nu}(t',\mbox{\bf r}') \rangle =
-i \delta_{\mu \nu}  \int \frac{d^{4} k}{(2 \pi)^{4}}
  \frac{\mbox{\Large $e^{i {\bf k}\cdot ({\bf r} -
  {\bf r}')}$ }
\mbox{\Large $e^{-i \omega (t - t')}$ }
}{ - \omega^{2} + \mbox{\boldmath $k$}^{2} - i \epsilon } \label{em}
\end{equation}
and the coupling to the 2 + 1 dimensional current is perfectly
consistent. In the computation of electronic properties we have
just to take care of placing the points ${\bf r} , {\bf r}'$ on the
plane of the two-dimensional layer. With this description
we certainly get more than we would need in a nonrelativistic theory,
but we can recover at any time the nonrelativistic limit of
all quantities by expanding
in powers of $v/c$.

It is important to stress that we have to expect, in general,
different answers in the computation of a given observable
taking (\ref{hcoul}) or (\ref{hq}) as
starting point.
The dimensionality of the system is low enough so
that the massless condition of the spinor reflects in the
appearance of infrared divergences in perturbation theory. Thus,
although we should not expect any discrepancy in the computation
of local quantities (like cutoff dependent quantities) in the
nonrelativistic model (\ref{hcoul})
 and in the model given by (\ref{hq}) in the
limit $v/c \rightarrow 0$, the response of nonlocal quantities
to the interaction turns out to be quite different in the
infrared regime. From a technical point of view, perturbation
theory looses predictive power because of the mentioned
infrared instabilities, and one has to resort to more
sophisticated methods in order to obtain information from the
quantum theory.

Let us illustrate the above fact with the computation of the quantum
corrections to the density of states $n(\omega )$ near the Fermi
level ($\omega = 0$). This observable is given by
\begin{equation}
 n(\omega ) = Im \;  \int d^{2} k \; Tr \: \left[ G(\omega,
   \mbox{\boldmath $k$}) \sigma_{3} \right]     \label{density}
\end{equation}
In the theory with the instantaneous Coulomb interaction,
the density of states is not modified
to first order in perturbation theory.
The instantaneous interaction can be obtained
from (\ref{em}), for instance, by considering formally its
expression for $c \rightarrow \infty$.
In this limit we have, computing always at
points ${\bf r}, {\bf r}'$ in the two-dimensional layer
\begin{eqnarray}
\langle T A_{\mu}(t,\mbox{\bf r}) A_{\nu}(t',\mbox{\bf r}') \rangle
  &  \approx &
-i \delta_{\mu \nu}  \int \frac{d^{2} k \;d \omega}{(2 \pi)^{3}}
  \int \frac{d k_{z}}{2 \pi}
  \frac{\mbox{\Large $e^{i {\bf k}\cdot ({\bf r} -
  {\bf r}')}$ }
\mbox{\Large $e^{-i \omega (t - t')}$ }
}{  \mbox{\boldmath $k$}^{2} + k_{z}^{2} - i \epsilon } \nonumber  \\
  & = &  - i \delta_{\mu \nu } \delta (t - t')
    \frac{1}{2} \int \frac{d^{2} k }{(2 \pi)^{2}} \frac{1}{ |
      \mbox{\boldmath $k$}| }
\mbox{\Large $e^{i {\bf k}\cdot ({\bf r} -
  {\bf r}')}$ }  \label{propa}
\end{eqnarray}
Then, it becomes obvious that this kind of interaction cannot
modify the $\omega$ dependence of
the propagator $G$, to the first perturbative order
(see also the discussion at the end of section 4).
However, as mentioned above, the propagator (\ref{propa}) cannot
be safely used in the computation of nonlocal quantities.  This remark is
appropriate to the case of the density of states, whose
determination in the
quantum field theory given by (\ref{hq}) goes as follows.

The inverse propagator may be  decomposed, as usual, into a
free part and
the electron self-energy $\Sigma (\omega, \mbox{\boldmath $k$})$
\begin{equation}
\frac{1}{G} = \frac{1}{G^{(0)} } - \Sigma
\end{equation}
$\Sigma (\omega, \mbox{\boldmath $k$})$ has a perturbative expansion in
terms of one-particle-irreducible diagrams\cite{landau},
which are in general
ultraviolet divergent. Since our model can be treated as a
genuine quantum field theory we expect these divergences to be
local and susceptible of being absorbed into $1/G^{(0)}$. For
the moment we are only interested in finite corrections and will
leave the issue of renormalization to the next section. To first
order in perturbation theory, the relevant diagram is shown
in figure 3 and gives (in the
limit $v/c \rightarrow 0$) the finite contribution to $\Sigma$
\begin{equation}
\Sigma_{R} (\omega, \mbox{\boldmath $k$}) =
 i \frac{e^{2}}{16 \pi^{2}}
v \; \mbox{\boldmath $\gamma \cdot k$} \; \int_{0}^{1} dx
\frac{\sqrt{ 1 - x }}{(1 - x + v^{2}x)^{2} }
log \left\{ v^{2} \mbox{\boldmath $k$}^{2} x - \omega^{2} x(1 - x) \right\}
 + O(e^{4})
\end{equation}
When inserted in equation (\ref{density}) the above expression
gives the first quantum corrections to the density of levels
\begin{eqnarray}
n(\omega)  &  =  &  2 \pi^{2} \frac{|\omega|}{v^{2}} \left(
 1 + \frac{1}{32 \pi} \frac{e^{2}}{v} \; log \;\omega^{2}
 + \frac{1}{8 \pi^{2}} e^{2} \int_{0}^{1} dx
\frac{\sqrt{ 1 - x }}{(1 - x + v^{2}x)^{2} } \; log \; x
\nonumber  \right.       \\
 &  & \left. + \frac{1}{16 \pi^{2}} e^{2} \int_{0}^{1} dx
\frac{\sqrt{ 1 - x }}{(1 - x + v^{2}x)^{2} } \frac{1}{x}
 \;\; + \;\; O(e^{4}) \right)
\label{21}
\end{eqnarray}
It is already appreciated that the perturbative expansion of
$n(\omega )$ is not well-defined, since the last integral in
(\ref{21}) is clearly divergent. Even if we could manage to
regularize in some way this infrared divergence, we still would
be left with a more serious problem. At the point near which we
want to measure the density of levels, i.e. near $\omega = 0$,
the first perturbative correction dominates over the zeroth order
term, and one can presume that higher order terms can produce
increasingly large powers of $log \: \omega$.
Perturbation theory is clearly
not reliable in the infrared regime, but the kind of infrared
instability described points at new physics
near the Fermi level.

\section{Renormalization}

Henceforth we will pay attention to the
local quantum field theory (\ref{hq}), which offers the
possibility of analyzing the infrared behaviour by application
of renormalization group methods.
We study first in this section the renormalization properties
of the model. Though the many-body theory of the 2D layer
has a natural ultraviolet cutoff, we insist in absorbing
ultraviolet divergent contributions into bare parameters of the
theory since this is a way of extracting
relevant information about the scaling properties in the
infrared\cite{zj,amit}.
As we have mentioned before, the renormalizability of a quantum
many-body theory is, in general, a nontrivial issue, since the
couplings in the effective theory may depend on the
point chosen on the Fermi surface. In our case things become much
simpler as our low-energy effective theory is a
genuine quantum field theory.
Regarding the usual considerations for renormalizability in
field theory, the only condition that we are missing is
relativistic invariance since the Fermi velocity of the
electrons $v = 3 ta/2 $ does not match the speed of light $c$
that appears in the dispersion relation of photons. For this
reason it is still necessary to check that the ultraviolet
divergences of the theory can be absorbed into a redefinition of
the scale of the fields and of the parameters in the hamiltonian.
We present here the analysis at the one-loop level, and conclude
with a compelling argument showing that the model is
renormalizable to all orders in perturbation theory.

The conditions which make the theory renormalizable can be
stated more clearly in the lagrangian formalism. Corresponding
to the hamiltonian (\ref{hq}), we have the action
\begin{eqnarray}
S  &  =  &   \int dt d^{2} r \; \overline{\Psi}( - \gamma_0
\partial_0 + v \mbox{\boldmath $\gamma \cdot \nabla$}) \Psi \nonumber  \\
  &    &  - i e  \int dt d^{2} r \; \overline{\Psi}( - \gamma_0
A_0 + v \mbox{\boldmath $\gamma \cdot A$}) \Psi
\end{eqnarray}
In the quantum theory the coefficients of the terms in the
action have to be adjusted so as to cancel the ultraviolet
divergences. We should start therefore with the bare action
\begin{eqnarray}
S_{bare}  &  =  & Z_{kin} \int dt d^{2} r \; \overline{\Psi}( - \gamma_0
\partial_0 + Z_{v} v \mbox{\boldmath $\gamma \cdot \nabla$})
  \Psi \nonumber  \\
  &    &  - Z_{int} \; i e  \int dt d^{2} r \; \overline{\Psi}( - \gamma_0
A_0 + Z_{v} v \mbox{\boldmath $\gamma \cdot A$}) \Psi   \label{action}
\end{eqnarray}
The renormalization coefficient $Z_{kin}$ may be determined from
the electron self-energy $\Sigma (\omega, \mbox{\boldmath $k$})$, as
well as the renormalization coefficient for the Fermi velocity $Z_{v}$.
The coefficient $Z_{int}$ is determined from the renormalization
of the time component of the interaction vertex. It turns out
therefore that the spatial components have necessarily to be
renormalized by the product $Z_{int} Z_{v}$. This is a
nontrivial check of the renormalizability of the theory, which
arises in the absence of relativistic invariance. Provided that
it is fulfilled, the renormalized theory may be made finite in
terms of renormalized quantities $t_{R}, e_{R}, \Psi_{R},
A_{R}^{\mu}$
\begin{eqnarray}
v_{bare}  &  =  &  Z_{v} v_{R}   \nonumber   \\
e_{bare}  &  =  &  Z_{e} e_{R}   \nonumber   \\
\Psi_{bare}  &  =  &  Z_{\Psi}^{1/2} \Psi_{R}   \nonumber   \\
A_{bare}^{\mu}  &  =  &  Z_{A}^{1/2} A_{R}^{\mu}  \label{ren}
\end{eqnarray}

In the previous section we already quoted the one-loop renormalized
contribution to the electron self-energy $\Sigma (\omega,
\mbox{\boldmath $k$})$. We deal here with the ultraviolet divergent part
of this object, regularizing it by working in analytical
continuation to dimension $d = 3 - \varepsilon$\cite{ramond}. From the
diagram in figure 3 we have
\begin{eqnarray}
\lefteqn{  \frac{1}{G^{(0)} } - \Sigma}  &    &  \nonumber \\
 &  =  &  - Z_{\Psi} i( -\gamma_{0} \omega + Z_{v} \; v
 \;  \mbox{\boldmath $\gamma \cdot k$} )
 - i \frac{e^{2}}{8 \pi^{2}} \; \gamma_0 \omega \;
 \left( 1 - 2 v^{2} \right) \;
\int_{0}^{1} dx \frac{\sqrt{ 1 - x }}{1 - x + v^{2}x }
 \frac{1}{\varepsilon} \nonumber  \\
 & &  - i \frac{e^{2}}{8 \pi^{2}} \;
\; v \; \mbox{\boldmath $\gamma \cdot k$} \;
\int_{0}^{1} dx \frac{\sqrt{ 1 - x }}{(1 - x + v^{2}x)^{2} }
\frac{1}{\varepsilon}
\; + \; finite \;\; terms \;\; + \;\; O(e^{4})
\end{eqnarray}
In order to ensure the finiteness of the electron propagator we
take the renormalization coefficients
\begin{eqnarray}
Z_{\Psi} & = & 1 + \frac{e^{2}}{8 \pi^{2}} \left( 1 - 2 v^{2} \right)
\int_{0}^{1} dx \frac{\sqrt{ 1 - x }}{1 - x + v^{2}x }
\; \frac{1}{\varepsilon} \;  + \; O(e^{4})  \label{zpsi}  \\
Z_{v} & = & 1 - \frac{e^{2}}{8 \pi^{2}}
\int_{0}^{1} dx \frac{\sqrt{ 1 - x }}{(1 - x + v^{2}x)^{2} }
\; \frac{1}{\varepsilon} \;    \nonumber      \\
  &   &  \;\;\;\;\; - \frac{e^{2}}{8 \pi^{2}} \left( 1 - 2 v^{2} \right)
\int_{0}^{1} dx \frac{\sqrt{ 1 - x }}{1 - x + v^{2}x }
\; \frac{1}{\varepsilon} \;  + \; O(e^{4})
\end{eqnarray}
These coefficients have a more complicated structure than those
of a relativistic theory, as long as $v$ does not match the speed
of light ---which, in our units, means that $v \neq 1$. $Z_{v}$
has, for instance, an infinite power series expansion in $v$
\begin{eqnarray}
Z_{v} & = & 1 - \frac{1}{16 \pi^{2}} e^{2}
 \left\{ \pi \frac{1}{v} F \left(\frac{1}{2}, \frac{3}{2}; \frac{1}{2};
 v^{2} \right) - 2 \pi v \left( 1 - 2 v^{2} \right)
   F \left(\frac{3}{2}, \frac{3}{2}; \frac{3}{2};
 v^{2} \right)   \right.  \nonumber   \\
      &   &    \left. + 4 \left( 1 - 2 v^{2} \right)
F \left(1, 1; \frac{1}{2}; v^{2} \right) -
4 F \left(1, 2; \frac{3}{2};
 v^{2} \right)  \right\} \frac{1}{\varepsilon}
  +  O   \left( e^{4}   \right)    \label{alt}
\end{eqnarray}
We will later use this expression in the consideration of the
nonrelativistic approximation $v \rightarrow 0$.

Now we come to the renormalization of the interaction term in
the action (\ref{action}). The first quantum corrections to the
vertex are given by the diagram in figure 4, and we denote them
symbolically by $\Gamma_{\mu}$. The divergent contribution to
the time component of the vertex is actually different than that
to the spatial components. We find that
\begin{eqnarray}
\Gamma_0  & = & i \; \gamma_0 \;
 \frac{e^{3}}{16 \pi^{2}}
  \left( 1 - 2 v^{2} \right)    \int_{0}^{1} dx
\frac{1 - x}{\sqrt{x} \left( v^{2} + (1 - v^{2})x \right) }
   \nonumber   \\
  &  &  - i \; \gamma_0 \;
 \frac{e^{3}}{8 \pi^{2}} v^{2} (1 - 2 v^{2})  \int_{0}^{1} dx
\frac{1 - x}{\sqrt{x} \left( v^{2} + (1 - v^{2})x \right)^{2} }
\frac{1}{\varepsilon}           \nonumber              \\
 &  & \;\;\;  + \; finite \;\; terms \;\; + O(e^{5})      \\
\mbox{\boldmath $\Gamma$} & = &  - i \frac{e^{3}}{16 \pi^{2}}
 v  \mbox{\boldmath $\gamma$}   \int_{0}^{1} dx
\frac{1 - x}{\sqrt{x} \left( v^{2} + (1 - v^{2})x \right) }
  \frac{1}{\varepsilon}
 \; + \; finite \;\; terms \;\; + O(e^{5})
\end{eqnarray}
We may think of $\Gamma_{\mu}$ as a correction to the bare
vertex in (\ref{action}). We have then
\begin{eqnarray}
Z_{int}  & = &  1 - \frac{e^{2}}{16 \pi^{2}}
 \left( 1 - 2 v^{2}  \right)  \int_{0}^{1} dx
\frac{1 - x}{\sqrt{x} \left( v^{2} + (1 - v^{2})x \right) }
\frac{1}{\varepsilon}    \nonumber  \\
   &  &  +  \frac{e^{2}}{8 \pi^{2}}
  v^{2} (1 - 2 v^{2}) \int_{0}^{1} dx
\frac{1 - x}{\sqrt{x} \left( v^{2} + (1 - v^{2})x \right)^{2} }
 \frac{1}{\varepsilon}  \; +  O(e^{4})   \label{uno}  \\
Z_{int} Z_{v}  & = &  1 -  \frac{e^{2}}{16 \pi^{2}} \int_{0}^{1} dx
\frac{1 - x}{\sqrt{x} \left( v^{2} + (1 - v^{2})x \right) }
  \frac{1}{\varepsilon} \;  + O(e^{4})   \label{dos}
\end{eqnarray}
By taking the quotient between (\ref{dos}) and (\ref{uno}), we
get an independent determination of the renormalization
coefficient $Z_{v}$
\begin{eqnarray}
Z_{v}  & = &  1 -  \frac{e^{2}}{16 \pi^{2}} \int_{0}^{1} dx
\frac{1 - x}{\sqrt{x} \left( v^{2} + (1 - v^{2})x \right) }
  \frac{1}{\varepsilon}                 \nonumber  \\
  &  & +  \frac{e^{2}}{16 \pi^{2}} \int_{0}^{1} dx
\frac{1 - x}{\sqrt{x} \left( v^{2} + (1 - v^{2})x \right) }
\left( 1 - 2 v^{2} - \frac{2 v^{2} (1 - 2 v^{2})}
{ v^{2} + (1 - v^{2})x } \right) \;
\frac{1}{\varepsilon} \; + O(e^{4})    \nonumber  \\
   & = &  1 - \frac{e^{2}}{16 \pi^{2}}
 \left\{ \pi \frac{1}{v} (1 - 2 v^{2})
F \left(\frac{1}{2}, \frac{1}{2}; - \frac{1}{2};
 v^{2} \right) + \frac{8}{3}  v^{2} \left( 1 - 2 v^{2} \right)
   F \left(2, 2; \frac{5}{2};
 v^{2} \right)   \right.  \nonumber   \\
      &   &    \left. + 2 \pi v
F \left(\frac{1}{2}, \frac{3}{2}; \frac{1}{2}; v^{2} \right) -
8 v^{2} F \left(1, 2; \frac{3}{2};
 v^{2} \right)  \right\} \frac{1}{\varepsilon}
  +  O   \left( e^{4}   \right)   \label{one}
\end{eqnarray}
It is not obvious that this expression coincides with the
alternative form (\ref{alt}), though it has to for the theory to
be renormalizable. In fact, after using the recursion relations
which hold for hypergeometric functions\cite{grad} one can show that the
two functions in (\ref{one}) and (\ref{alt}) are identical. We
then have the nice situation of dealing with a
nonrelativistic quantum field theory in which the Fermi velocity
of the electrons is renormalized exactly in the same way in the
different terms of the action.

This precise coincidence is not casual. One
is led to suspect that it has to be consequence of some symmetry
operating in the theory, which turns out to be nothing but gauge
invariance. Actually, this should automatically lead in the
quantum theory to the same renormalization of the kinetic part
and the interaction term in the action, $Z_{kin} = Z_{int}$.
It is not difficult to check that this is satisfied in our model,
to the perturbative order in which we are working. From
expressions (\ref{zpsi}) and (\ref{uno}) we have
\begin{eqnarray}
Z_{kin}  & = &  1 + \frac{e^{2}}{8 \pi^{2}} (1 - 2 v^{2})
   \left\{ 2 F \left(1, 1 ;  \frac{1}{2};
 v^{2} \right) - \pi v  F \left(\frac{3}{2}, \frac{3}{2}; \frac{3}{2};
 v^{2} \right)  \right\}  \frac{1}{\varepsilon}
  +  O   \left( e^{4}   \right)       \label{zkin}    \\
Z_{int}  & = &   1 - \frac{e^{2}}{16 \pi^{2}}
  (1 - 2 v^{2}) \left\{ - \frac{\pi}{v}
F \left(\frac{1}{2}, \frac{1}{2}; - \frac{1}{2};
 v^{2} \right) - \frac{8}{3}  v^{2}  F \left(2, 2; \frac{5}{2};
 v^{2} \right)   \right.  \nonumber   \\
      &   &    \left. +  \frac{\pi}{v}
F \left(\frac{1}{2}, \frac{3}{2}; \frac{1}{2}; v^{2} \right) -
4  F \left(1, 2; \frac{3}{2};
 v^{2} \right)  \right\} \frac{1}{\varepsilon}
  +  O   \left( e^{4}   \right)
\end{eqnarray}
which become identical after use again of the relations among
hypergeometric functions.

The gauge symmetry
\begin{eqnarray}
 \Psi  &  \rightarrow  &  \mbox{\Large $e^{i e \theta (t, {\bf
  r}) }$} \Psi                   \nonumber    \\
 A_{\mu}  & \rightarrow &  A_{\mu} + \partial_{\mu} \theta (t, {\bf
  r})
\end{eqnarray}
is manifest at the classical level. Provided that it holds
for the bare action (\ref{action}), it ensures that
$Z_{kin} = Z_{int}$. It is easily seen, by taking a gauge
parameter $\theta$ independent of $t$, that it also
enforces the same renormalization of
the Fermi velocity $v$ in the different terms of the action. As
long as one is able to preserve the gauge invariance in the
quantum theory, this turns out to guarantee the
renormalizability to all orders of the perturbative expansion.

The rest of the renormalization coefficients in (\ref{ren}) can
be obtained from the following remark. The fact that
$Z_{\Psi}$ has to equal $Z_{int}$ to all orders in perturbation
theory implies, as in standard quantum electrodynamics,
that $Z_{e} Z_{A}^{1/2} = 1$. The renormalization coefficient
of the electromagnetic field $Z_{A}^{1/2}$ is determined from
the polarization tensor, whose first
perturbative contribution shown diagrammatically in figure 5 is a
finite quantity
\begin{equation}
\Pi_{\mu \nu}(k) = \frac{1}{2 \pi} \frac{e^{2}}{v^{2}}
  (g_{\mu \nu} - \frac{k_{\mu}
k_{\nu} }{k^{2}}) |k| \int_{0}^{1} dx \: \sqrt{x (1 - x)}
\label{pol}
\end{equation}
We conclude, therefore, that
\begin{equation}
Z_{A}^{1/2} = Z_{e}^{-1} = 1 + O \left( e^{4} \right) \label{charge}
\end{equation}
remarking the absence of renormalization of the electric charge
to first order in perturbation theory.

\section{Scaling of the electronic interaction (I).
Nonrelativistic regime.}

In this section we seek the connection of our model to the
real physics of electrons in a two-dimensional layer. A single
graphite sheet or any fullerene aggregate of sufficiently large
size is the appropriate physical instance to look at.
The first question concerns the magnitude of the parameters
pertinent to the real layer. A standard value for the
hopping parameter in graphite is $t \approx 2.2 eV$ . For the nearest
neighbor separation in the lattice we can take $a \approx 0.14
\cdot 10^{-9} m$ . From these values and restoring for a moment
the dependence on $c$ and $\hbar$, the Fermi velocity in our
model turns out to be $v =  3 ta/(2 \hbar) \approx 1.5 \cdot
10^{-3} c$ . This is certainly much smaller than the speed of
light, for which to extract relevant quantities to the electron
physics in the layer we will have to take the limit $v/c \rightarrow 0$
and keep the leading order in previous formulas.

For a clearer description of our approach to renormalization
group we will quote the renormalization coefficients in this
section as functions of the ultraviolet momentum cutoff
$\Lambda$ rather than in terms of the dimensional regularization
poles $1/\varepsilon $. At the one-loop level this just amounts
to replace $1/\varepsilon $ by $log \; \Lambda$ in all formulas.
The renormalization coefficient of the Fermi velocity $Z_v$
becomes, for instance, in the nonrelativistic limit
\begin{equation}
Z_v = 1 - \frac{1}{16 \pi} \frac{e^{2}}{v} log \; \Lambda +
O\left( \frac{e^{4}}{v^{2}} \right)   \label{zetav}
\end{equation}
It is not difficult to see that at each level of the
perturbative expansion the leading order in the nonrelativistic
approximation enters with a power $( e^{2}/v )^{n} $. We define,
therefore, a consistent perturbative expansion in the
nonrelativistic limit by taking $v \rightarrow 0$ (or,
equivalently, $c \rightarrow \infty$) and $e^{2}/v =
const. $ This latter quantity becomes then the effective
strength of the electronic interaction. In this limit we obtain
from (\ref{zkin})
\begin{equation}
Z_{\Psi} = 1 + O\left( \frac{e^{4}}{v^{2}} \right)
\end{equation}
so that we do not find wavefunction renormalization at the
one-loop level in the nonrelativistic limit. We recall also the
former result (\ref{charge})
\begin{equation}
Z_{A}^{1/2} = Z_{e}^{-1} = 1 + O \left( e^{4} \right)
\end{equation}

We have a nonrelativistic quantum theory with no renormalization
of the electric charge, but in which the effective strength of
the interaction $e^{2}/(4 \pi v)$ is renormalized. We rely on
the original interpretation of renormalization adopted
by Wilson\cite{wilson}
to determine the scale dependence of the effective interaction.
Since we have shown that the model is renormalizable, we may
impose the independence of the renormalized theory on the
momentum cutoff $\Lambda$. By means of renormalization group
transformations we are able therefore
to relate bare theories at different
scales (different cutoff) which correspond to the same
renormalized theory\cite{zj,amit}.
The essential information is encoded in the
renormalization group equation, which for the electron
propagator, for instance, states that under a change in the
scale of the cutoff $\Lambda$
\begin{equation}
\left( \Lambda \frac{\partial }{\partial \Lambda} +
 \beta_{v}(v, e^{2}) \frac{\partial }{\partial v}  +
 \beta_{e}(v, e^{2}) \frac{\partial }{\partial e}  -
         \gamma (v, e^{2})
   \right)  G(\omega , \mbox{\boldmath $k$} , \Lambda ;
   v, e^{2})  =  0
\end{equation}
where
\begin{eqnarray}
 \beta_{v} (v, e^{2})  & = &  \Lambda \frac{\partial Z_{v}
  }{\partial \Lambda } v_{R}       \label{betav}  \\
 \beta_{e} (v, e^{2})  & = &  \Lambda \frac{\partial Z_{e}
  }{\partial \Lambda } e_{R}                 \\
 \gamma  (v, e^{2})  & = &  \Lambda \frac{\partial
  }{\partial \Lambda }  \; log \: Z_{\Psi}
\end{eqnarray}
The well-known solution to this equation takes the form
\begin{equation}
  G(\omega, \mbox{\boldmath $k$} , \rho \Lambda ; v, e^{2} )  =
   \exp{ \left\{ \int^{\rho }  \frac{d \rho '}{\rho '} \; \gamma
\right\} }
\; \; G (\omega, \mbox{\boldmath $k$} , \Lambda ;  v_{eff}(\rho),
  e^{2}_{eff}(\rho) )           \label{rge}
\end{equation}
where the effective parameters are given by
\begin{eqnarray}
\rho \frac{\partial }{\partial \rho} v_{eff} (\rho ) & = &
 - \beta_{v} ( v_{eff},  e^{2}_{eff} )         \nonumber   \\
 \rho \frac{\partial }{\partial \rho} e_{eff} (\rho ) & = &
 - \beta_{e} ( v_{eff},  e^{2}_{eff})
\end{eqnarray}

Increasing the scale $\rho $ of the cutoff is equivalent to
measure the observables of the theory
at large distance scale. From the one-loop order
results of the previous section we see
that $e_{eff}$ is constant at this level, while $v_{eff}$
is not. We expect, therefore, the effective coupling of the
electronic interaction $e^{2}/(4 \pi v)$ to have a nontrivial
renormalization group flow in the infrared regime. Actually,
from equations (\ref{zetav}) and (\ref{betav})
\begin{equation}
\rho \frac{\partial }{\partial \rho} \frac{v_{eff}}{v_{R}}
= \frac{1}{16 \pi} \frac{e^{2}}{v_{eff}}
+  O \left( \frac{e^{4}}{v^{2}_{eff}} \right) \label{flow}
\end{equation}
so that the asymptotic behaviour of the coupling is
\begin{equation}
\frac{1}{4 \pi} \frac{e^{2}}{v_{eff}}(\Lambda / \Lambda_{0}) =
\frac{1}{4 \pi} \frac{e^{2}}{v_{R}}
 \left( \frac{v_{0}^{2}}{v_{R}^{2}} + \frac{1}{8 \pi}
 \frac{e^{2}}{v_{R}}
\; log \frac{\Lambda}{\Lambda_{0}} \right)^{-1/2} \label{sol}
\end{equation}

We have to bear in mind that, when integrating equation
(\ref{flow}), we are relying heavily on perturbation theory.
If we recall our original estimate of the Fermi velocity $v$, we
see that the coupling $e^{2}/(4 \pi v)$ of the nonrelativistic
theory is not a small parameter but, in fact, larger than one.
Nevertheless, we know that when dealing with a renormalized quantum theory
the statement that a coupling like $e^{2}/(4 \pi v)$ is
small or large applies to a particular energy scale. We then have to
be very precise in establishing the regime in
which the integration of the renormalization group equation
becomes consistent. Let us remark that the solution (\ref{sol})
of the renormalization group equation is equivalent to
performing the sum of leading logarithm terms of the
perturbative expansion and curing, therefore, the problem of the
infrared instability of the density of states
addressed at the end of section 2. Taking the limit $\Lambda
\rightarrow \infty$  in the bare theory corresponds to measure
at large distances (or
small energies with respect to the Fermi level) and in this
regime perturbation theory becomes more and more accurate.
This is another instance of a
physical system which has strong coupling at certain energies
and an asymptotic regime at which perturbation theory becomes
reliable, similar to what is believed to be true (as supported
by experiment) in the quantum field theory of the strong
interactions. We are providing the example of a nonrelativistic
theory in which this also happens, exchanging the ultraviolet
regime in the case of the strong interactions by the infrared
regime here.

As for the practical application of the flow (\ref{sol}),
we remark that the nonrelativistic theory
appears to be formally incomplete, since it does not give any
hint about what the value of $v_{R}$ has to be. This is related
to the unlimited growth of the Fermi velocity, and to the fact
that, in the nonrelativistic approximation, there is no signal of
a renormalization group fixed point. Later we will see that in the
complete quantum field theory the fixed point is given by
$v_{eff} = c$, but the flow in the graphite layer may not be
necessarily controlled by this value of the renormalized
coupling.
We believe, anyhow, that the observation of the phenomenon
of scaling in the asymptotic region should be possible in a
single graphite sheet.

Regarding the possible existence of a natural infrared cutoff,
it can be shown that in our model
the Coulomb interaction is not screened,
in the usual metallic manner, by
quantum effects. This is due to the vanishing of
the density of states at the Fermi level. Thus, a finite
screening length is only generated by giving to the system an
excess of charge, as a consequence of doping.
In order to study this effect, we may place the system away
from half-filling by introducing a chemical potential $\mu $ for
the conserved charge
\begin{equation}
\frac{n}{V} = - \lim_{t' \rightarrow t + 0} Tr \; \langle T \Psi
(t, {\bf 0}) \Psi^{\dag} (t', {\bf 0})  \rangle  \label{ncharge}
\end{equation}
The correct expression of the free electron propagator at finite
$\mu $ is\cite{chin}
\begin{equation}
  \left. G^{(0)} (\omega, \mbox{\boldmath $k$}) \right|_{\mu \neq 0} =
  i \frac{ - \gamma_{0} \omega
+ v \; \mbox{\boldmath $\gamma \cdot k$}}
{ - \omega^{2} + v^{2} \mbox{\boldmath $k$}^{2} - i \epsilon }
  - \pi \frac{ - \gamma_{0} \omega
+ v \; \mbox{\boldmath $\gamma \cdot k$}}
{ v \left| \mbox{\boldmath $k$} \right| } \theta (\mu - v \left|
\mbox{\boldmath $k$}
 \right|) \delta (\omega - v \left| \mbox{\boldmath $k$} \right|)
\end{equation}
Obviously, the $\mu$-dependent part cannot change the
ultraviolet behaviour of the loop-integrals and the
renormalization coefficients of the quantum field theory are not
modified by the nonvanishing chemical potential. This may lead,
however, to finite corrections. The fermion loop diagram is now
nonvanishing, opposite to what happened before at $\mu = 0$. The
computation of (\ref{ncharge}) gives
\begin{equation}
\frac{n}{V} = \frac{1}{(2 \pi)^{2}} \int d^{2} k \: d \omega
  \frac{\omega}{ v | \mbox{\boldmath $k$} |} \theta (\mu - v |
       \mbox{\boldmath $k$} |)
   \delta (\omega - v | \mbox{\boldmath $k$} |) = \frac{1}{4 \pi}
   \frac{\mu^{2}}{v^{2}}
\end{equation}
which, after multiplying by 2 (the two independent Fermi points)
gives the correct number of particles.
Regarding the polarization tensor, a nonvanishing value of
$\Pi_{0 0}(\omega = 0, \mbox{\boldmath $k$} \rightarrow
 {\bf 0})$ signals the
appearance of a photon effective mass. At small $\mu $ one
finds actually the estimate for the inverse of the screening length
\begin{equation}
  \Pi_{0 0}(0, \mbox{\boldmath $k$} \rightarrow {\bf 0})
  \approx  \frac{e^{2}}{2 \pi } \frac{\mu }{v^{2}}
\end{equation}
which gives a measure of the screening effects away from half-filling.

We now turn, finally, to the issue of whether the electron
observables may get anomalous dimensions in the nonrelativistic
limit. We remark that the absence of one-loop  wavefunction
renormalization in the limit $v \rightarrow 0$ is just an accident.
The signal of wavefunction renormalization is given by an
ultraviolet divergent self-energy contribution of the form
\begin{equation}
\Sigma (\omega, \mbox{\boldmath $k$}) \approx  f \left(
\frac{e^{2}}{v} \right) \gamma_0 \omega  +  \ldots
\end{equation}
To establish the degree of divergence of the local contributions
one can safely implement the nonrelativistic limit in the photon
propagator, which leads to the expression (\ref{propa}).
The $\omega$-dependence in the vertex
diagram in figure 3 can be absorbed then by a simple change
of variables in the
loop integral, so that the divergent part of the diagram cannot
bear any dependence on it.
At the two-loop level, the same independence of the local
contributions on $\omega$ holds except for the vertex
diagrams in figure 6. In these diagrams the loop integrals
overlap in such a way that the previous argument does not apply.
By using the nonrelativistic approximation (\ref{propa}) and
after a very lengthy calculation, one obtains from the sum of the
two-loop diagrams
\begin{equation}
   Z_{\Psi}  =  1 - \frac{ c_{1}}{ 16 \pi^{2} }
               \frac{e^{4}}{v^{2}} log \Lambda
    \;\; + \;\; O(e^{6})
\end{equation}
where $c_{1}$ is a nonvanishing quantity, $c_{1} \approx 5.49
\cdot 10^{-2}$. We see therefore that the
electron observables acquire anomalous dimensions in the
infrared regime. According to equation (\ref{rge}), the electron
propagator transforms under a change of scale $\rho $ by the
factor
\begin{equation}
   \exp{ \left\{ - \frac{ c_{1} }{16 \pi^{2} }
    \int^{\rho }  \frac{d \rho '}{\rho '} \;
   \frac{e^{4}}{v^{2}_{eff}} (\rho ') \right\} }
\end{equation}

We have therefore an electronic system which deviates slightly
from Fermi liquid behaviour. In particular, we have seen that the
vanishing of the density of states at the Fermi level suppresses
the usual screening of the Coulomb interaction. This property is
maintained in the quantum theory, at least at the one-loop
level, since the Fermi level is not shifted by quantum corrections.
Also a gap does not open at the Fermi points to the one-loop order
(as it would be if an effective mass for the spinor had been
generated in perturbation theory). The system is not an
insulator, though it does not exhibit usual metallic properties either.

\section{Scaling of the electronic interaction (II). Non-Fermi
liquid fixed point.}

In the previous section we have mentioned that the
nonrelativistic theory is formally incomplete since it does not
know of any bound in the effective Fermi velocity $v_{eff}$ of
electrons. As this parameter grows steadily at large distances
the nonrelativistic approximation becomes less accurate, up to a
point in which the power series expansion in $v/c$ looses its
meaning. In this relativistic regime, we have to use the full
expression of the beta function $ \beta_{v} (v, e^{2})$ to
analyze the behaviour of $v_{eff}$. In particular, we are
interested in knowing if a possible fixed point of the
renormalization group flow exists, characterized by the condition
\begin{equation}
 \beta_{v} (v, e^{2})  = 0
\end{equation}
By quoting (\ref{alt}) in terms of elementary functions, this
equation reads, at the one-loop level,
\begin{equation}
\frac{1}{v} \frac{1 - 2 v^{2} + 4 v^{4}}{ (1 - v^{2})^{3/2} } \;
arccos \: v  \; + \;  \frac{ 1 - 4 v^{2} }{ 1 - v^{2} }  =  0
\label{fix}
\end{equation}
For positive $v$, the only solution to (\ref{fix}) is $v =
1$ \footnote{This is, obviously, a statement regarding the limit
value of the left hand side of (\ref{fix}) as $v \rightarrow 1$.},
which in our units means $v$ equal to $c$. Thus, the full
quantum field theory has a built-in upper bound of the renormalization
group flow, and the speed of light turns out to be the fixed
point for the Fermi velocity of electrons.

In fact, it is not difficult to check that the one-loop beta
function has a simple zero at $v = 1$, and to compute the flow
in the neighborhood of the fixed point. The renormalization
group equation for $v_{eff}$ becomes
\begin{equation}
\rho \frac{\partial }{\partial \rho}  \frac{ v_{eff} }{ v_{R} }
 \approx  \frac{2}{5} \frac{ e^{2} }{ \pi^{2} } ( 1 - v_{eff} )
\end{equation}
The asymptotic behaviour of the solution is, taking for $v_{R}$
the fixed point value,
\begin{equation}
log ( 1 - v_{eff} ) \approx  - \frac{2}{5} \frac{ e^{2} }{
  \pi^{2} } log \left( \Lambda / \Lambda_{0} \right)
\end{equation}
In figure 7 we show the plot of this scaling behaviour,
which is practically indistinguishable in the range considered
from that obtained with the nonrelativistic approximation.

Recalling the vanishing of the electric charge beta function
$ \beta_{e} (v, e^{2}) $ at the one-loop level, we may assert
that to this perturbative order we have identified an infrared
fixed point given by $v_{eff} = c$ and arbitrary (though small)
value of $e_{eff}$. The one-loop result is not satisfactory,
however, since the independence of $e_{eff}$ on the flow may be
destabilized by higher-loop effects. In the
renormalization group approach to Fermi liquid theory\cite{shan}, the
consideration of the one-loop graph in figure 5 is enough since
it encodes, via the renormalization group, information of the
series of ladder diagrams. In the usual analysis of the many-body
theory with an extended Fermi surface these are which dominate
in the infrared regime. Because of the existence of Fermi points
in the honeycomb lattice at half-filling, though, there is no
argument here to conclude the relevance of one set of diagrams
against the rest. In our quantum field theory description it
seems that no diagram can be discarded a priori. It is also for
this reason that there are diagrams which are irrelevant in
Fermi liquid theory but give rise in our model to wavefunction
renormalization and anomalous dimensions in the infrared regime.

Regarding the renormalization of the electric charge, we have
analyzed the polarization tensor $\Pi_{\mu \nu}$ to second order
in perturbation theory. As stated above, the renormalization of
the photon field $A_{\mu }$ is directly related by gauge
invariance to the renormalization of the charge $e$. Computing
in the framework of dimensional regularization we have not found any
divergent contribution of order $e^{4}$ to $\Pi_{\mu \nu}$. The
reason is the same operating in the one-loop polarization tensor.
A na\"{\i}ve dimensional analysis of the diagram shows that it should
diverge linearly in momentum space. In the dimensional
regularization scheme this leads to the appearance of gamma
functions at half-integer values of the argument, instead of
poles at $d \rightarrow 3$. We believe that this scheme is the
best suited for the regularization of the gauge theory ---it
certainly is not going to break the gauge invariance in any event.
If, however, one insists on understanding the regularization in
terms of a cutoff dependence, it can be checked that this does
not lead either to a renormalization  of the scale of the field
$A_{\mu}$. One would have to look for divergent contributions of
$\; log \: \Lambda$ type accompanying the gauge invariant
expression (\ref{pol}) for $\Pi_{\mu \nu}$. These are
not present at the two-loop order and one may conjecture,
following the above dimensional argument, that they do not appear
to any order in perturbation theory.

At or near the fixed point, the two-dimensional
electron liquid we are describing is definitely not a Fermi liquid.
The main reason for this assertion is the anomalous dimension
acquired by the electron field $\Psi $, which may be obtained
from (\ref{zpsi}) evaluated now at $v = 1$. At the fixed point
the behaviour of the Green function under a change of scale is
\begin{equation}
\left. G(\omega, \mbox{\boldmath $k$}) \right|_{\Lambda}
 \equiv \Lambda^{\gamma}
\Phi (\omega, \mbox{\boldmath $k$})
\end{equation}
or, equivalently, after introducing the precise value of
$\gamma$
\begin{equation}
G( \rho \omega, \rho \mbox{\boldmath $k$} ) =
  \rho^{-1 + e^{2}/(12 \pi^{2})}
G( \omega, \mbox{\boldmath $k$} )
\end{equation}
We arrive in this way to solve the problem of the proliferation
of $\; log \: \omega$ terms in the perturbative computation of the
density of states. As is well-known, the renormalization group
approach succeeds in summing up
all these increasingly divergent contributions, in the
form of an anomalous dependence on the electron frequency. We
have found that this sum is consistently achieved in the
relativistic regime, where the fixed point lies and the electron
field has a simple, anomalous scaling behaviour.

\section{Summary and conclusions}

We have studied a system of electrons in the two-dimensional
honeycomb lattice interacting via a Coulomb potential.
The fact that the Fermi surface of this lattice at half-filling
consists of two isolated points renders the problem similar
to the $d=1$ case and allows the formulation of the
low energy effective theory as a standard
quantum field theory. We have been able to investigate the
infrared behavior of the system by
using renormalization group
techniques and
have seen that, indeed,
it resembles more that of one-dimensional electrons
than that of a two-dimensional Fermi liquid.

Before entering into the analysis of the phenomenological
consequences of our work, there are a
few points that we would like to emphasize.

Let us first comment on the
use made through the paper of the electromagnetic field
propagator.
This use is unconventional both from the condensed matter
point of view where the photon is mostly absent,
and from the quantum field theory since, as we will
strength now,
we are not dealing with
standard massless QED.

Expression (\ref{em}) shows the usual three-dimensional photon
of QED. This is what we use in our relativistic
computations. The constraint that,
in the electronic interactions, the two points
lay on the lattice
plane, allows us to integrate out the perpendicular dimension.
The resulting expression read as a propagator on the plane
certainly differs from the standard propagator
in QED$_3$. In particular its infrared behavior is
quite different, it goes like $1/r$ instead of
the  standard $log \; r$. This is the origin of, for
instance, the finiteness of the photon self-energy up to
the two-loop level and, hence, of the nonrenormalization
of the electric charge.

Let us now comment on the comparison between
the instantaneous versus the retarded propagator.
The effect of retardation on
the infrared behavior of nonlocal
quantities has been exemplified with the computation
of the quantum corrections to the density of
states at the end of section 2.
In the theory given by the Coulomb interaction (\ref{hcoul}),
no one-loop corrections were found
to the free value of the density of states.
In the theory with the retarded interaction
our RG analysis shows that,
despite the infrared divergence found at the one-loop level,
the approach is consistent in the sense that the
perturbative expansion of the nonrelativistic approximation
does not have to make sense
in the infrared when the Fermi velocity is still far from the
fixed-point value. Hence,
we must resort to the full relativistic theory
when computing the infrared behavior of nonlocal
observables. There, the RG allows to interpret the one-loop
infrared instability shown in section 2 as the manifestation, in
perturbation theory, of the anomalous dimension.

We have then found that our original system flows
in the infrared towards
a fixed point of the renormalization group that is a
relativistic quantum field theory with anomalous dimension of
the fermionic propagator.

Our analysis has important consequences in
condensed matter physics.
One of the more interesting is the anomalous screening
of the Coulomb interactions of the model.
In a Fermi liquid, formation of virtual electron-hole pairs
(i. e., quantum corrections to the photon
self-energy)
generates a new scale, the screening
length.
This is the basis of the absence of long-range
interactions in the ultralocal approach to
interacting electrons (Hubbard model).
What we have seen is that, due to the vanishing
of the density of states at the Fermi level
and to the absence  of renormalization of the photon
self-energy, there
is no metallic screening in our materials \cite{app}. The
robustness of this result stands on the non-renormalization of
the chemical potential (the Fermi level remains a set of two points after
the corrections due to the interactions).
Moreover we have seen that, although the
low energy density of states goes to zero, no
finite gap appears in the spectrum of charge excitations.
Our results rule out the validity of descriptions
in terms of on-site interactions only. Screening
effects are not strong enough to reduce the couplings
to simple on-site terms.

The one-particle properties are very different from
those of a conventional Fermi system. The wavefunction
renormalization leads to the disappearance of the
quasiparticle peak at low energies. Thus, no effective
description in terms of independent, and coherent,
quasiparticles is possible. The situation resembles
the 1D Luttinger liquid, where any amount of
electron-electron interaction suffices to wipe out
the quasiparticle pole, leading to incoherent
one-particle excitations.

Finally, the existence of nontrivial scaling laws
implies that anomalous exponents should show up in
a variety of properties, like the specific heat,
susceptibility, etc. The possibility of having
an experimental test of these results remains open.

To conclude this summary and to point out possible future work,
we would like to remark the interest of
investigating the connections of
our work with the infrared catastrophe mentioned in the
introduction as well as with other unconventional
Fermi liquid behavior discussed in the literature
\cite{nf}. On this respect, this work can be seen as a rigorous
realization of the ideas settled in \cite{varma}.

\vspace{1cm}

{\large \bf Acknowledgements}

This work has been partially supported by the CICYT.

\newpage
\section*{Figure Captions}
\mbox{  }

{\bf Figure 1:} The planar honeycomb lattice.

{\bf Figure 2:} Representation in $(E,{\bf k})$ space of the lower
branch of the electronic dispersion relation (in units $t = a = 1$).
The cusps appear at the six corners of the first Brillouin zone.

{\bf Figure 3:} One-loop contribution to the electron self-energy.

{\bf Figure 4:} First quantum corrections to the interaction vertex.

{\bf Figure 5:} One-loop polarization tensor.

{\bf Figure 6:} Two-loop diagrams contributing to wavefunction
renormalization.

{\bf Figure 7:} Plot of the effective Fermi velocity $ v_{eff} $
(in units $c = 1$) versus length scale $L/L_{0}$.

\end{document}